\documentclass[pdftex,twocolumn,epjc3]{svjour3}          

\RequirePackage[T1]{fontenc}

\smartqed  
\pdfoutput=1
\RequirePackage{graphicx}
\usepackage{mathtools}
\usepackage{float}
\usepackage{siunitx}
\RequirePackage{mathptmx}      
\RequirePackage{flushend}
\RequirePackage{enumerate}
\RequirePackage[numbers,sort&compress]{natbib}
\RequirePackage[colorlinks,citecolor=blue,urlcolor=blue,linkcolor=blue]{hyperref}
\RequirePackage{lineno}
\makeatletter

\journalname{Eur. Phys. J. C}

\begin{document}

\title{Study of charged hadron multiplicities  in  charged-current neutrino-lead interactions in the OPERA detector}

\author{
N.~Agafonova\thanksref{addr1}
\and
A.~Aleksandrov\thanksref{addr2}
\and
A.~Anokhina\thanksref{addr3}
\and
S.~Aoki\thanksref{a4}
\and
A.~Ariga\thanksref{a5}
\and
T.~Ariga\thanksref{a5,a6}
\and
A.~Bertolin\thanksref{a7}
\and
I.~Bodnarchuk\thanksref{a8}
\and
C.~Bozza\thanksref{a9}
\and
R.~Brugnera\thanksref{a7,a10}
% ...
}
\thankstext[$\star$]{c1}{Corresponding authors. E-mail: ali.murat.guler@cern.ch, Cagin.Gunes@ankara.edu.tr}

\institute{
INR - Institute for Nuclear Research of the Russian Academy of Sciences, RUS-117312 Moscow, Russia\label{addr1}
\and
INFN Sezione di Napoli, 80125 Napoli, Italy\label{addr2}
\and
SINP MSU - Skobeltsyn Institute of Nuclear Physics, Lomonosov Moscow State University, RUS-119991 Moscow, Russia\label{addr3}
\and
 Kobe University, J-657-8501 Kobe, Japan\label{a4}
 \and
Albert Einstein Center for Fundamental Physics, Laboratory for High Energy Physics (LHEP), University of Bern, CH-3012 Bern, Switzerland\label{a5}
\and
  Faculty of Arts and Science, Kyushu University, Japan\label{a6}
  \and
 INFN Sezione di Padova, I-35131 Padova, Italy\label{a7}
  \and
 JINR - Joint Institute for Nuclear Research, RUS-141980 Dubna, Russia\label{a8}
  \and
 Dipartimento di Fisica dell'Universit\`a di Salerno and ``Gruppo Collegato'' INFN, I-84084 Fisciano (Salerno), Italy\label{a9}
  \and
 Dipartimento di Fisica e Astronomia dell'Universit\`a di Padova, I-35131 Padova, Italy\label{a10}
% ...
}
   
\date{Received: date / Accepted: date}

\maketitle

\begin{abstract}

The OPERA experiment was designed to 
search for  $\nu_{\mu} \rightarrow \nu_{\tau}$  oscillations  in appearance mode
through the direct observation of tau neutrinos in the CNGS neutrino beam.
In this paper, we report a study of the multiplicity of charged particles produced
 in charged-current neutrino interactions in lead.
We present charged hadron average multiplicities, their dispersion and
investigate the KNO scaling in different kinematical regions.
 The results are presented in detail in the form of tables that can be used in the validation of Monte Carlo generators of neutrino-lead interactions.
\end{abstract}

\section{Introduction}
The multiplicity distribution of charged hadrons is an important characteristic of the hadronic final
 states in hard scattering processes. Since it reflects the dynamics of the interaction, it has been extensively studied in cosmic rays, fixed target and 
collider experiments \cite{Biebl,PLUTO,TASSO,Allen,CHORUS}.
These data are useful to improve models of particle production which are used in  Monte Carlo (MC) 
event generators. 

In this paper, we report the result on charged hadron production initiated in 
charged-current \mbox{$\nu_{\mu}$}  interactions in the OPERA target. The basic unit that constitutes the target is the Emulsion Cloud Chamber (ECC) detector 
which is a stack of  nuclear emulsion films 
acting as high precision trackers  interleaved with  lead plates that provide a massive target.
The excellent spatial resolution of nuclear emulsion allows the determination of the event topology
and  the measurement of charged particle trajectories.
Therefore, it is well suited for  the  investigation of
the moments of the  charged particles multiplicity distribution.
 However, only few studies of charged particle 
multiplicity in 
neutrino-nucleon  interactions were made using the nuclear emulsion technology 
\cite{CHORUS,Voyvodik,Aleshin}. 

In the following, a short description of the experimental setup and of the procedure used to locate neutrino interactions 
in the target is given; the data sample and the analysis procedure are presented. Then, multiplicity moments and investigation of 
KNO scaling \cite{Koba} 
in different kinematical regions are presented in a form suitable for  the validation 
of MC generators of neutrino$-$lead interactions.

\section{Experimental Procedure}

The OPERA experiment has been designed for the observation of $\nu_{\mu} \rightarrow \nu_{\tau}$  oscillations in  appearance mode in the CNGS 
(CERN Neutrinos to Gran Sasso) 
neutrino beam. 
The detector was located at the underground Gran Sasso Laboratory (LNGS) in  
Italy. It was exposed to the CNGS  muon neutrino beam with mean energy  17 \mbox{GeV} at a distance of \num{730} \mbox{km}.
OPERA reported the discovery of $\nu_\tau$ appearance 
with a significance of 5.1$\sigma$ \cite{OPERAprl}.
The OPERA detector is a hybrid setup consisting of electronic detectors and a massive lead-emulsion target segmented into 
ECC units, called bricks. The detector is composed of two identical Super-Modules, each of which has a target
section followed by a muon spectrometer which is composed of 
a dipole magnet instrumented  with resistive plate chambers and drift tubes. 
Each target section has 31 brick walls 
interleaved with orthogonal pairs of scintillator strip planes that compose the Target Tracker (TT).
A detailed description of the OPERA detector can be found in \cite{operadet}.

\begin{figure}[!ht]
  \centering
  \includegraphics[scale=0.42]{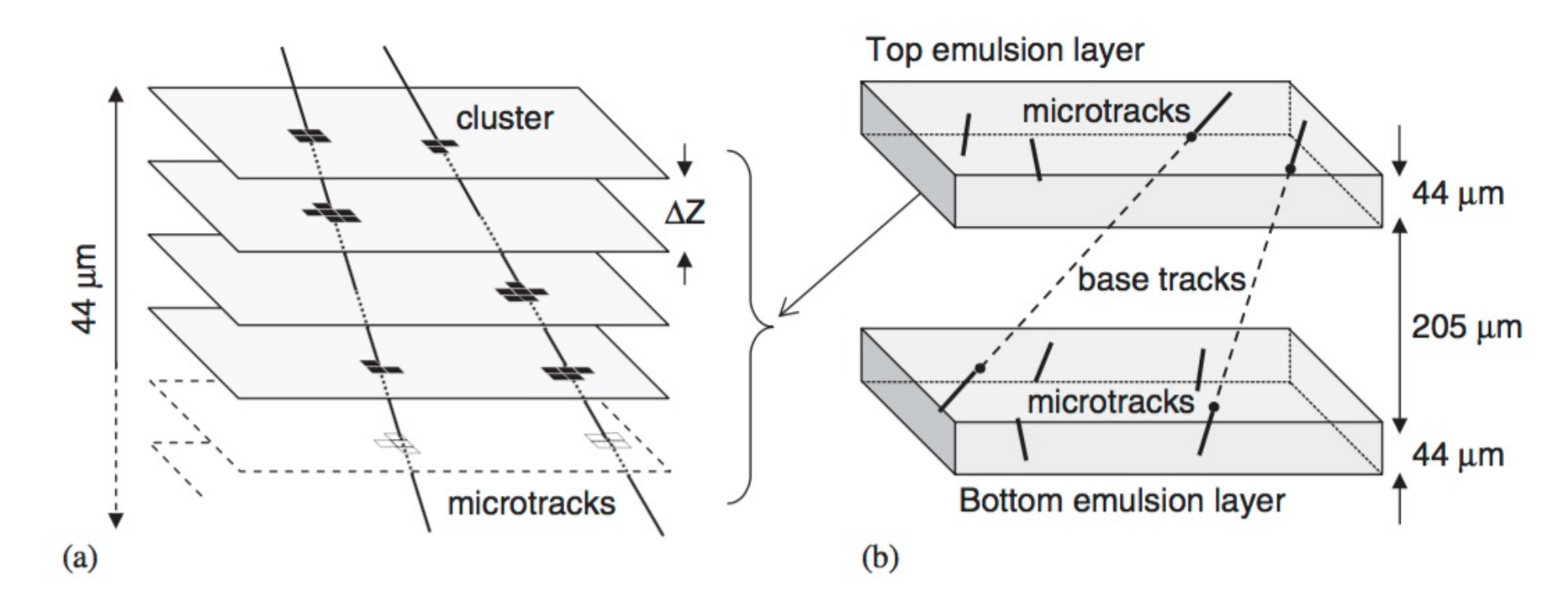}
  \caption{(a) Combination of clusters belonging to images taken at different depths in
one emulsion layer forms a micro-track; (b) The association of two micro-tracks across the plastic base forms a base-track.}\label{fig:micro}
\end{figure}
A brick  consists of 57 emulsion
films interleaved with 56 lead plates of 1 mm thickness.  The  films  are made of 2 emulsion layers, each
44 \mbox{$\mu$m} thick  coated on both sides of a 205 \mbox{$\mu$m} transparent plastic base. 
The brick has transverse dimensions of $128\times 103$ $\mbox{mm}^2$, a thickness of 81 mm (about $10$ radiation lengths) and
it weighs 8.3 kg. A pair of removable emulsion films called changeable sheets (CS) attached to the downstream face of each brick  
act as interfaces between the emulsion films in the brick and the TT.
There are about \num{150000} bricks in total for a target mass of 1.25 ktons. 

TT hit patterns are used to identify the bricks possibly containing the neutrino interaction vertex \cite{yuri}. The most probable brick is 
then extracted from the target and its CS films scanned. If a signal compatible with the
TT predictions is found, the brick is disassembled and its films analysed.
Once the vertex has been located, a surrounding volume of about 2 $\mbox{cm}^3$ is scanned to determine the event topology.
Otherwise, the procedure is repeated in the next brick in the probability ranking.
Track recognition in an emulsion layer is based on  16 tomographic images taken by the sensor of an automated microscope
and equally spaced through the 44-\mbox{$\mu$m}  depth of the layer.
A sequence of aligned grains in a layer forms a micro-track (Figure~\ref{fig:micro}a) and the association of two 
matching 
micro-tracks on each side of the plastic base in a film constitutes a base-track (Figure~\ref{fig:micro}b). 
Track positions and slopes are determined by 
a linear fit 
through base-tracks in the analysed volume.
The details of the event analysis procedure are described in \cite{OPERAloc}.
\section{Analysis}

During the physics runs between 2008 and 2012,
OPERA collected data corresponding to
$1.8 $x$ 10^{20}$ protons on target. The electronic detectors recorded \num{19505} neutrino interactions in the target fiducial volume. 
The search of the neutrino vertex in the first and second most probable bricks plus some additional 
selections (see \cite{OPERAloc} for more details) resulted in a sample of \num{5603} located events out of which \num{4406} have 
an identified muon. 
For the present measurement  an unbiased sub-sample of \num{818} events occurring in the lead with a 
negatively charged muon identified by the muon spectrometer was selected in
order to  measure the track and vertex parameters in the target including a detailed check of the nuclear break-up and evaporation processes.

It is further required that $W^2$, the square of the invariant mass of the hadronic system measured with the electronic detector, 
is larger than 1 $\mbox{GeV}^2/\mbox{c}^4$ in order to eliminate the quasi-elastic contribution.
The final data sample contains 795 events with an identified muon.
The  contamination  from  $\nu_\mu$ neutral-current interactions in the final data sample is estimated by  MC simulations to be less than $1\%$.

Selected  $\nu_\mu$ CC  events are inspected carefully 
and tracks are classified as being left by a minimum ionisation particle (mip), grey and black depending on their ionisation features.
The mip tracks are left by the muon and by highly relativistic charged hadrons resulting from the cascade of interactions generated inside the target nucleus 
by the primary hadrons emitted with the muon at the neutrino-nucleon interaction.
Black tracks are produced by low energy fragments (protons, deuterons, alpha particles and heavier fragments) emitted from the excited target nucleus.
Black tracks are classified as backward or forward 
based on the emission direction.  
Grey tracks are left by slow particles which are interpreted as being recoil nucleons emitted during the nuclear cascade \cite{Powell}.

\begin{figure}[H]
  \centering
    \includegraphics[scale=0.45]{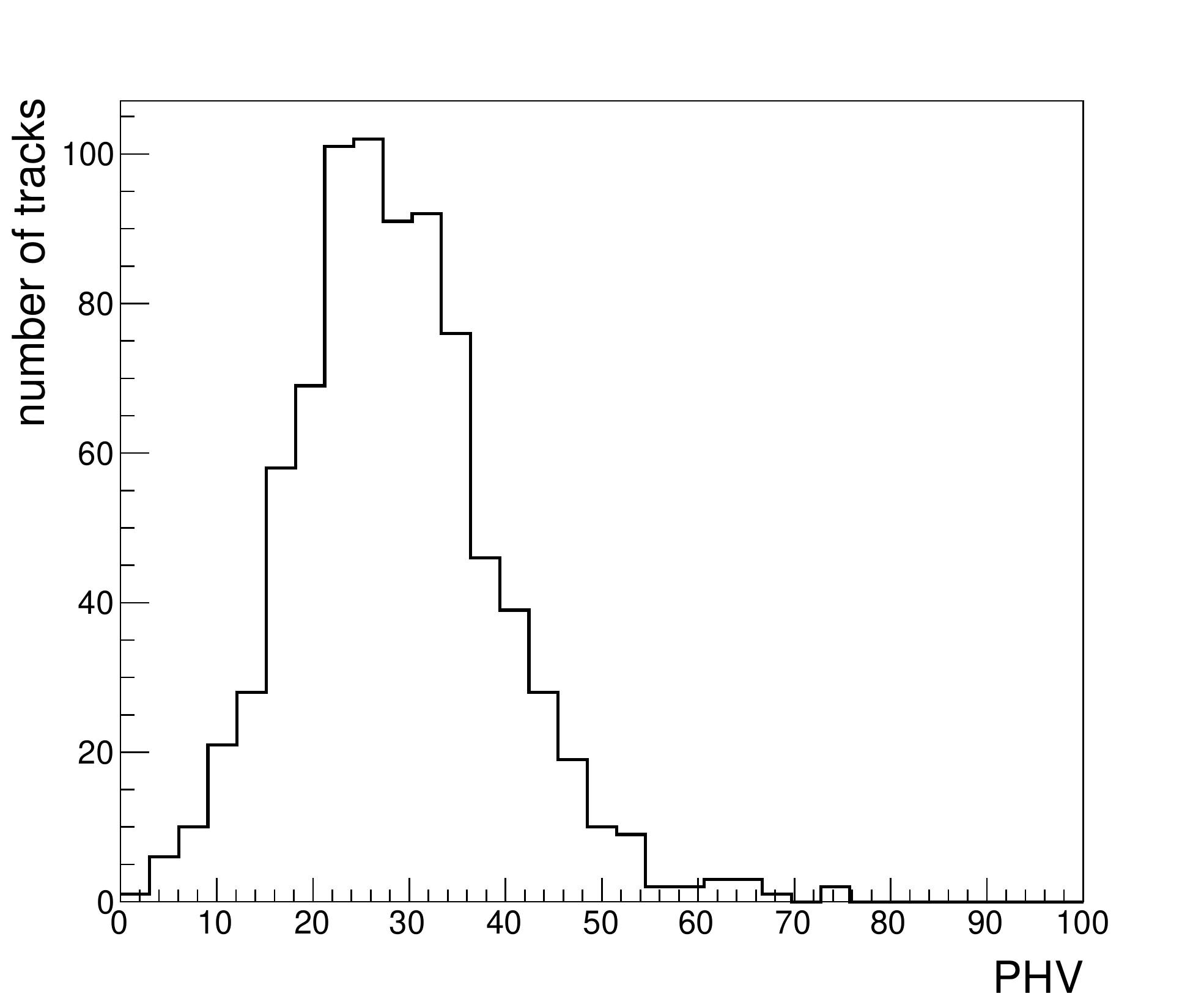}
  \caption{Pulse Height Volume distribution of muon tracks.}\label{fig:muonPHV}
\end{figure}

\begin{figure}[H]
  \centering
   \includegraphics[scale=0.45]{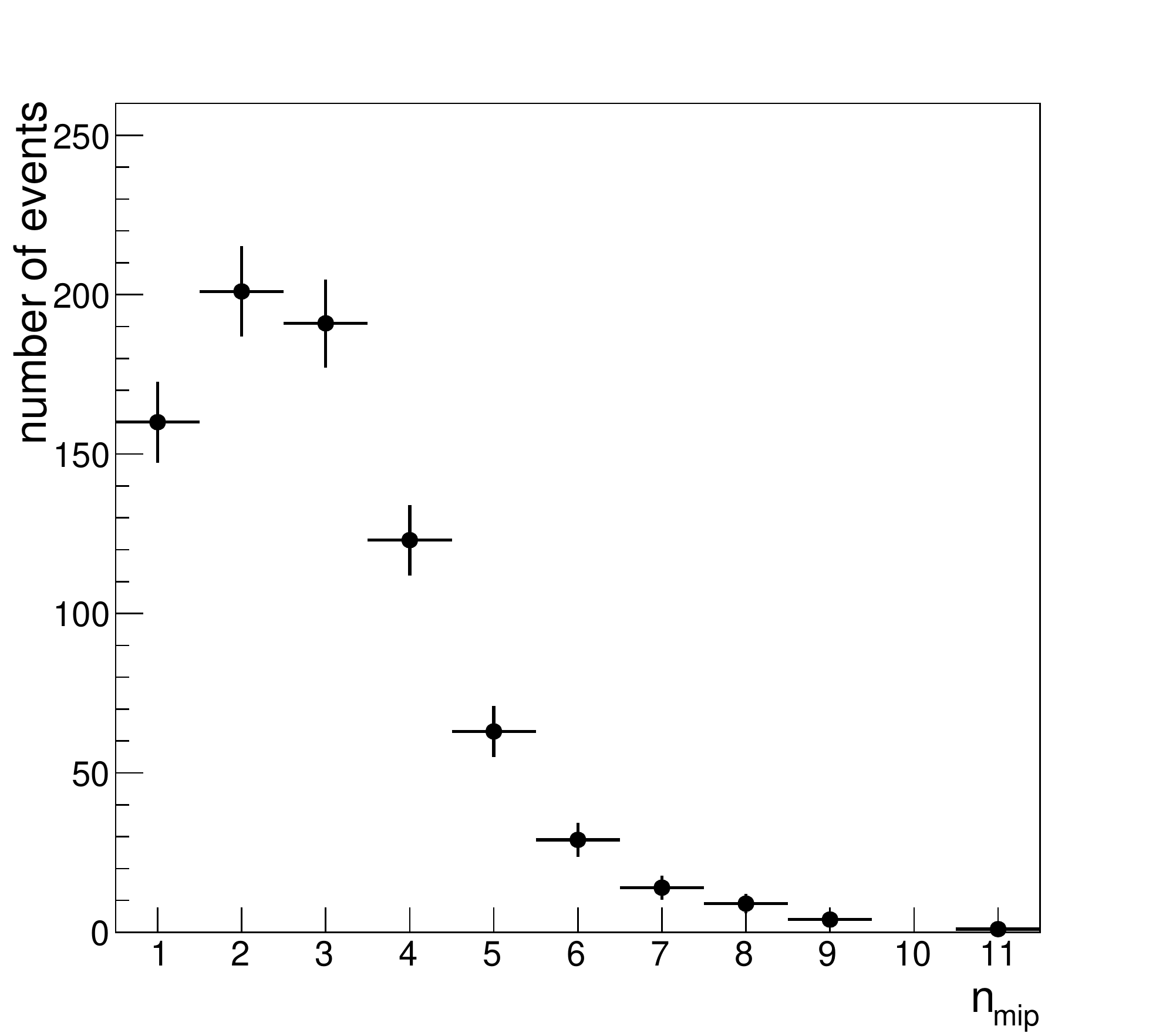}
 \caption{Multiplicity distribution of mip tracks.}\label{fig:multsh}
\end{figure}

\begin{figure}[H]
  \centering
    \includegraphics[scale=0.45]{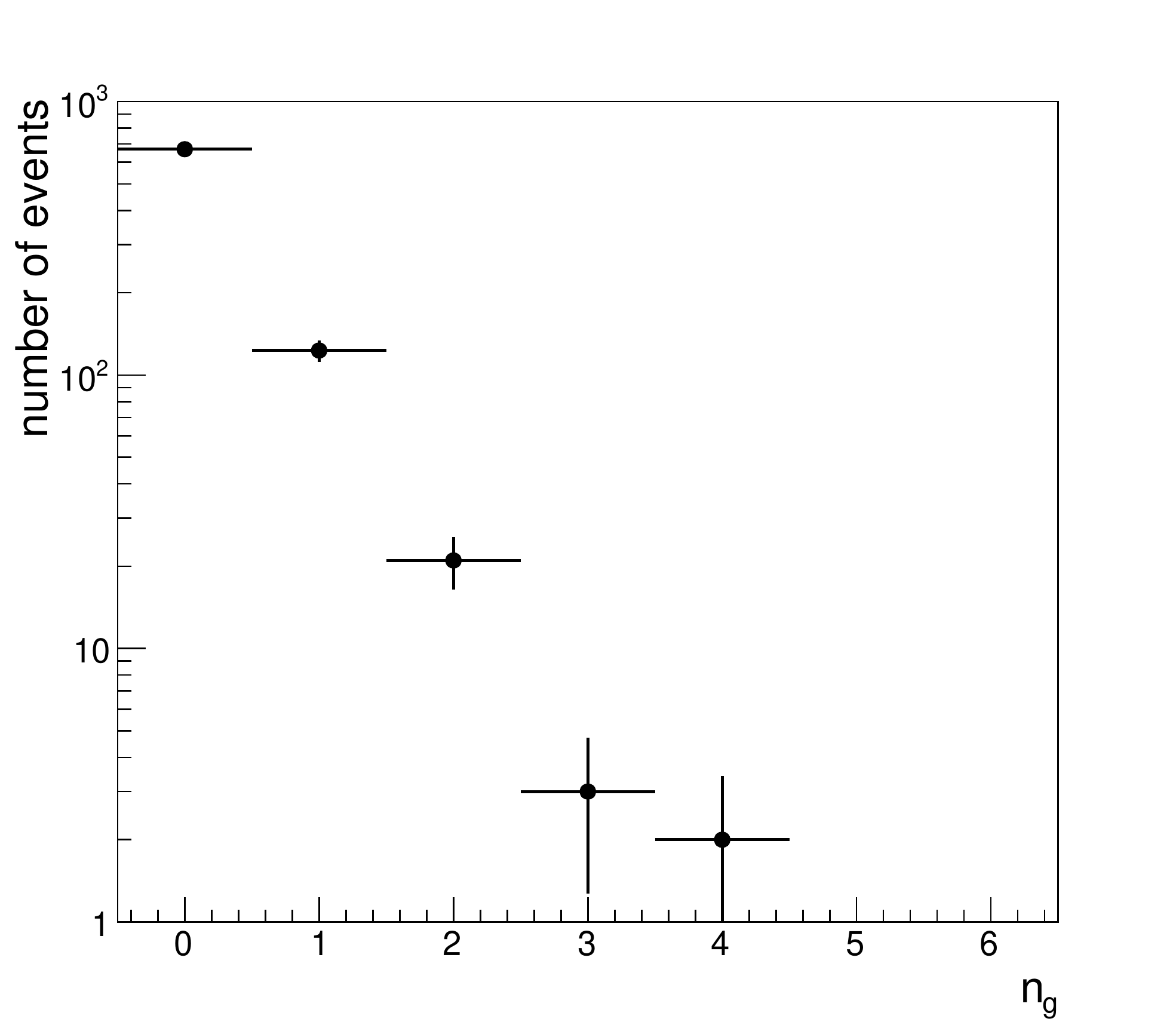}
  \caption{Multiplicity distribution of grey tracks.}\label{fig:multgrey}
\end{figure}

\begin{figure}[H]
  \centering
    \includegraphics[scale=0.45]{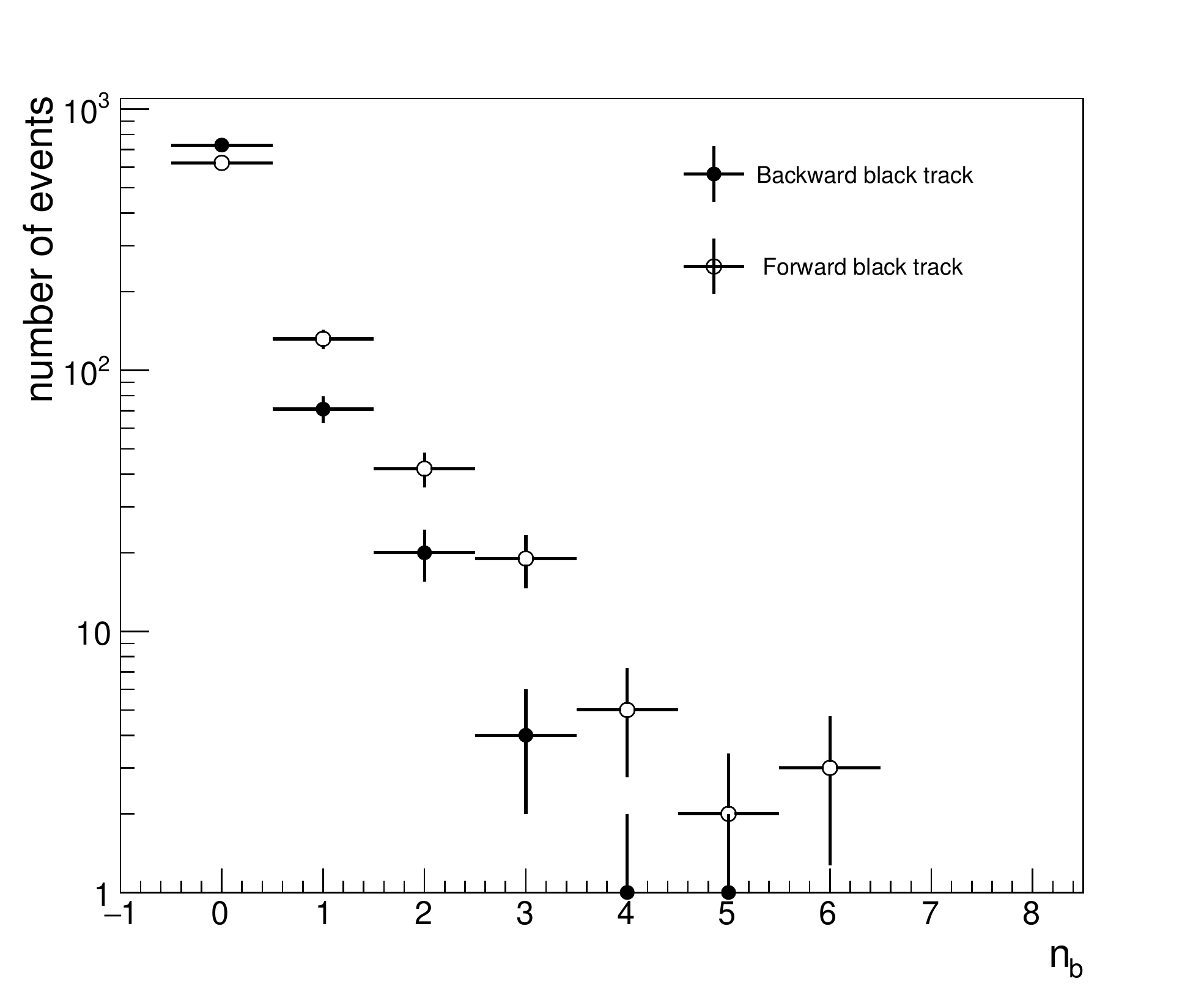}
     \caption{Multiplicity distribution of backward and forward black tracks.}\label{fig:multblack}
\end{figure}

The black tracks are easy to recognise visually since they are heavily ionising, they have short path lengths and stop within two lead plates.
The separation between mip and grey tracks is based on the 
 Pulse Height Volume (PHV) \cite{Toshito,Shin} which is defined as the sum of the number of
pixels associated with each track  in all sixteen layers of images \cite{ccd}. 
PHV indicates  the track width and is therefore  a  
measure of  the grain density of a track that reflects the energy deposition of a particle in the emulsion film. 
The PHV distribution of muon tracks is shown in Figure~\ref{fig:muonPHV}.   All muon tracks have a PHV  below 85 and mip tracks are defined by
requiring a PHV smaller than 85, the tracks with a higher PHV being classified as grey. 
The mip, grey and black track multiplicities are shown in Figures~\ref{fig:multsh}, \ref{fig:multgrey} and \ref{fig:multblack}, respectively.
%
The average numbers of mip and grey tracks in $\nu_\mu{CC}$ events are
$\langle n_{mip} \rangle = 2.94\pm 0.05$ and $\langle n_{g} \rangle = 0.22\pm 0.01$, respectively. 
The average number of backward and forward black tracks are measured to be
${\langle n_{b}\rangle}_B = 0.15\pm 0.01$ and ${\langle n_{b}\rangle}_F = 0.38\pm 0.02$, respectively.
The charged hadrons multiplicity $n_{ch}$ is defined as the number of mip tracks
excluding the muon track, its average being ${\langle n_{ch} \rangle} = {\langle n_{mip} \rangle} - 1 =
1.94\pm{0.05}$. The distribution of mip tracks as a function of their emission
angle with respect to the beam axis is given separately for muons and
charged hadrons in Table~\ref{tab:emiss}.
\begin{table}[h]
\centering
\addtolength{\tabcolsep}{-4pt}
\caption{The numbers of hadron ($n_{ch}$) and muon ($n_{\mu}$) tracks per event as a function of the emission angle $\theta$. }\label{tab:emiss}
\label{parset}
\begin{tabular*}{\columnwidth}{@{\extracolsep{\fill}}llll@{}}
\hline
\multicolumn{0}{@{}l}{ $\theta$(\mbox{Radian})} & {$\langle \theta \rangle$} &  $n_{ch}$/event & $n_\mu$/event \\
\hline
 0.00 $\div$ 0.050   & 0.031 $\pm$ 0.001   &0.134&   0.070\\

 0.050 $\div$ 0.100 &   0.076 $\pm$  0.001 & 0.258   &0.195\\

 0.100 $\div$ 0.150  &0.126 $\pm$  0.001& 0.280&  0.231\\

 0.150 $\div$ 0.200   &0.174 $\pm$  0.001 & 0.237 &  0.177\\

 0.200 $\div$ 0.300 &   0.246 $\pm$  0.002  & 0.383&0.195\\

 0.300 $\div$ 0.400 & 0.347 $\pm$  0.003  & 0.298 &0.079\\

 0.400 $\div$ 0.500 & 0.450 $\pm$  0.004  & 0.179&0.029\\

 0.500 $\div$ 0.600 & 0.549 $\pm$  0.006  & 0.113&0.017\\

 $\ge$  0.600 & 0.661 $\pm$  0.03   & 0.058&0.002\\
\hline
Total & ~ & $1.94$ & $1.00$\\
\end{tabular*}
\end{table}
\subsection{Efficiency Estimation}

Reconstruction and location efficiencies are computed using the standard OPERA simulation framework. 
The neutrino fluxes and spectra are based 
on the FLUKA  simulation \cite {fluka} of the CNGS beam-line. The neutrino interactions in the detector are generated 
using the NEGN generator \cite{Autiero}.
MC-generated $\nu_\mu CC$ events are processed through 
the full OPERA simulation chain, from the event 
classification and brick finding provided by the electronics detectors to the CS analysis and event location 
and analysis in the brick.

\begin{figure}[!ht]
\centering
    \includegraphics[scale=0.41]{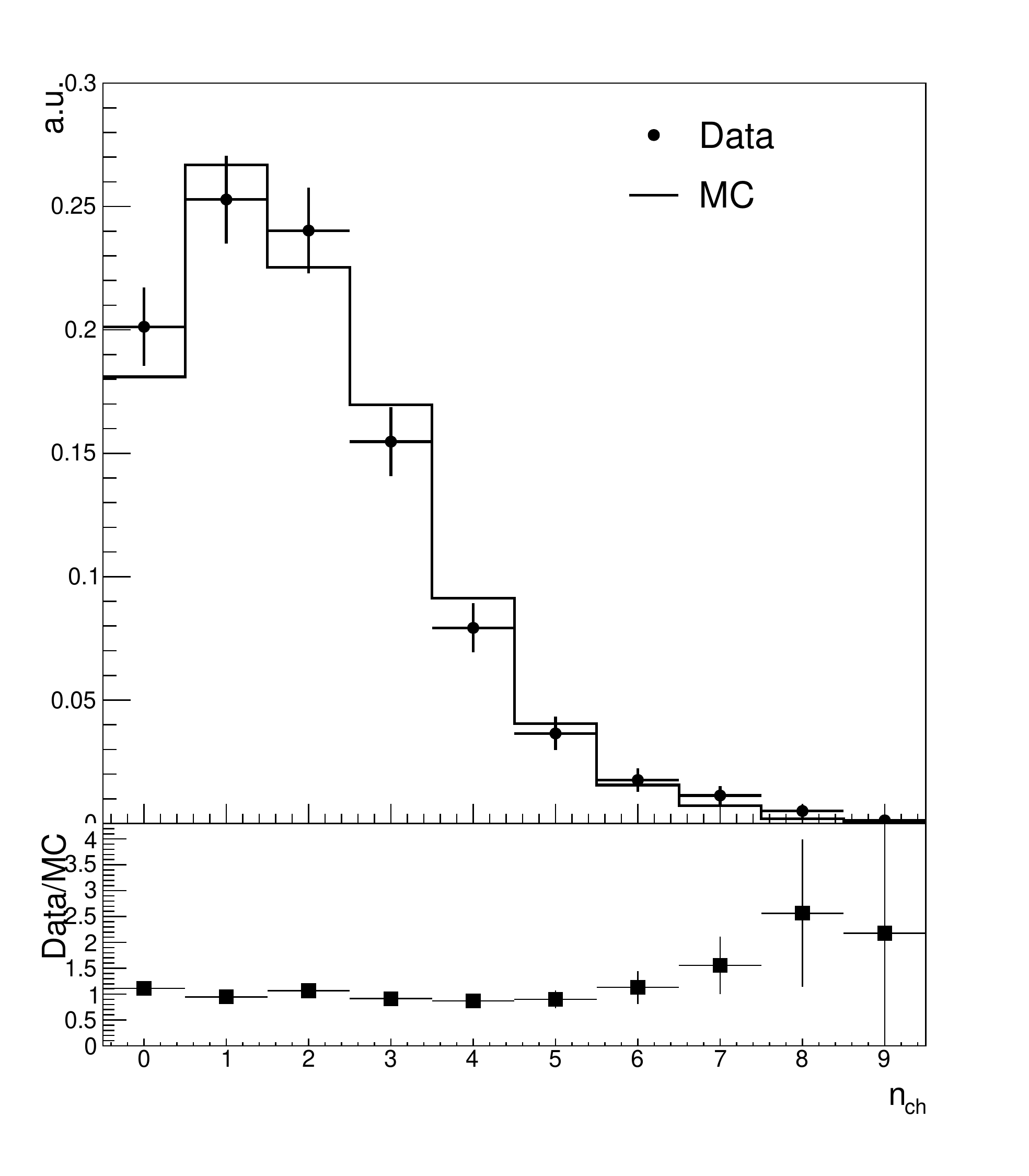}
  \caption{Data-MC comparison of the charged hadron multiplicity.}\label{fig:datamc}
\end{figure}

The location efficiency, shown in Table~\ref{tab:loc}, of the  $\nu_\mu CC$ events is estimated 
as a function of $W^2$ and of the charged hadron multiplicity.
Since the event location is done using mip tracks, the location efficiency does not depend 
on the black and grey track multiplicities at the neutrino interaction vertex. 
Figure~\ref{fig:datamc} shows the good agreement between the charged hadrons multiplicities obtained for observed and MC simulated data.
The location efficiency correction is hereafter 
applied to the measured data distributions. 

In $\nu_{\mu}CC$ interactions, neglecting the Fermi motion, the full kinematics of the event can be reconstructed from the measurement of the muon
momentum $p_{\mu}$, its angle $\theta_{\mu}$ with respect to
the beam axis, and $E_{had}$, the energy transfer to the hadronic system, obtained by measuring the energy deposition in the TT 
(detailed information is available in \cite{operadet})
\begin{align*}
E_{\nu}&=E_{\mu} + E_{had}\\
{Q^{2}_{\nu}}& = 2E_{\nu}(E_{\mu} - p_{\mu}\cos\theta_{\mu}  ) - m_{\mu}^{2}\\
W^{2} &= 2m_{N}E_{had} + m_{N}^{2} - Q_{\nu}^{2}\\\nonumber
\end{align*}
where $E_{\nu}$ and $E_{\mu}$ are the energy of the incoming neutrino and the muon, $Q_{\nu}^{2}$  
the squared four-momentum transfer, $m_{N}$
and $m_{\mu}$ are the mass of the nucleon
and muon respectively.
The average neutrino energy of the data sample is
$\langle E_{\nu}  \rangle$ = (19.6 $\pm$ 0.3) \mbox{GeV}, the mean-square momentum transferred to the hadronic system is  $\langle Q^{2}_{\nu}\rangle$ =  
(5.7 $\pm$ 0.3) $\mbox{GeV}^{2}$/$\mbox{c}^{2}$, and the mean-square of the invariant mass of the hadronic system is 
$\langle W^{2}\rangle$= (16.9 $\pm$ 0.6) $\mbox{GeV}^{2}/\mbox{c}^{4}$.
\begin{table*}
\caption{Location efficiency as a function of $\mbox{W}^2$ and the charged hadron $n_{ch}$ multiplicity.}\label{tab:loc}
\label{sphericcase}
\begin{tabular*}{\textwidth}{@{\extracolsep{\fill}}lrrrrrrl@{}}
\hline
{$W^{2}(\mbox{GeV}^{2}/\mbox{c}^{4}$)}& \multicolumn{1}{c}{$$} & \multicolumn{1}{c}{$$} & \multicolumn{1}{c}{$$} & \multicolumn{1}{c}{$n_{ch}$} & \multicolumn{1}{c}{$$} &  \multicolumn{1}{c}{$$}  &  \multicolumn{1}{c}{$$}  \\
\hline
{$$}& \multicolumn{1}{c}{$0$} & \multicolumn{1}{c}{$1$} & \multicolumn{1}{c}{$2$} & \multicolumn{1}{c}{$3$} & \multicolumn{1}{c}{$4$} &  \multicolumn{1}{c}{$\ge 5$}  &  \multicolumn{1}{c}{Total}  \\
\hline
1-3 & 0.32 $\pm$ 0.03 & 0.48 $\pm$ 0.02 & 0.53 $\pm$ 0.04  & 0.55 $\pm$ 0.06 & 0.56 $\pm$ 0.09 & 0.62 $\pm$ 0.17 & 0.46 $\pm$ 0.01  \\
3-6 &0.34 $\pm$ 0.02 & 0.47 $\pm$ 0.02 & 0.54 $\pm$ 0.03  & 0.56 $\pm$ 0.04 & 0.65 $\pm$ 0.07 & 0.66 $\pm$ 0.11 & 0.47 $\pm$ 0.01 \\
 6-9 & 0.33 $\pm$ 0.02 & 0.46 $\pm$ 0.02 & 0.56 $\pm$ 0.02  & 0.63 $\pm$ 0.03 & 0.67 $\pm$ 0.06 & 0.68 $\pm$ 0.08 & 0.48 $\pm$ 0.01  \\
9-12 & 0.34 $\pm$ 0.02 & 0.49 $\pm$ 0.02 & 0.57 $\pm$ 0.02  & 0.62 $\pm$ 0.03 & 0.69 $\pm$ 0.05 & 0.66 $\pm$ 0.07 & 0.50 $\pm$ 0.01  \\
12-15 & 0.36 $\pm$ 0.02 & 0.50 $\pm$ 0.02 & 0.58 $\pm$ 0.02  & 0.63 $\pm$ 0.03 & 0.70 $\pm$ 0.04 & 0.70 $\pm$ 0.06 & 0.52 $\pm$ 0.01  \\
15-19 &0.37 $\pm$ 0.02 & 0.49 $\pm$ 0.02 & 0.60 $\pm$ 0.02  & 0.62 $\pm$ 0.02 & 0.66 $\pm$ 0.03 & 0.74 $\pm$ 0.05 & 0.53 $\pm$ 0.01 \\
 19-25 &0.40 $\pm$ 0.02 & 0.52 $\pm$ 0.02 & 0.56 $\pm$ 0.02  & 0.64 $\pm$ 0.02 & 0.66 $\pm$ 0.03 & 0.68 $\pm$ 0.04 & 0.54 $\pm$ 0.01 \\
25-35 & 0.41 $\pm$ 0.02 & 0.56 $\pm$ 0.02 & 0.56 $\pm$ 0.02  & 0.60 $\pm$ 0.02 & 0.62 $\pm$ 0.02 & 0.69 $\pm$ 0.03 & 0.55 $\pm$ 0.01 \\
 $\ge$35   & 0.39 $\pm$ 0.02 & 0.39 $\pm$ 0.02 & 0.42 $\pm$ 0.01  & 0.43 $\pm$ 0.01 & 0.54 $\pm$ 0.02 & 0.57 $\pm$ 0.02 & 0.45 $\pm$ 0.01 
\\\hline
Total  & 0.37 $\pm$ 0.007 & 0.48 $\pm$ 0.007 & 0.54 $\pm$ 0.008  & 0.57 $\pm$ 0.09 & 0.62 $\pm$ 0.01 & 0.62 $\pm$ 0.01 & 0.50 $\pm$ 0.01
\end{tabular*}
\end{table*}
\subsection{Multiplicity Distributions}

\begin{figure}[h]
\centering
    \includegraphics[scale=0.45]{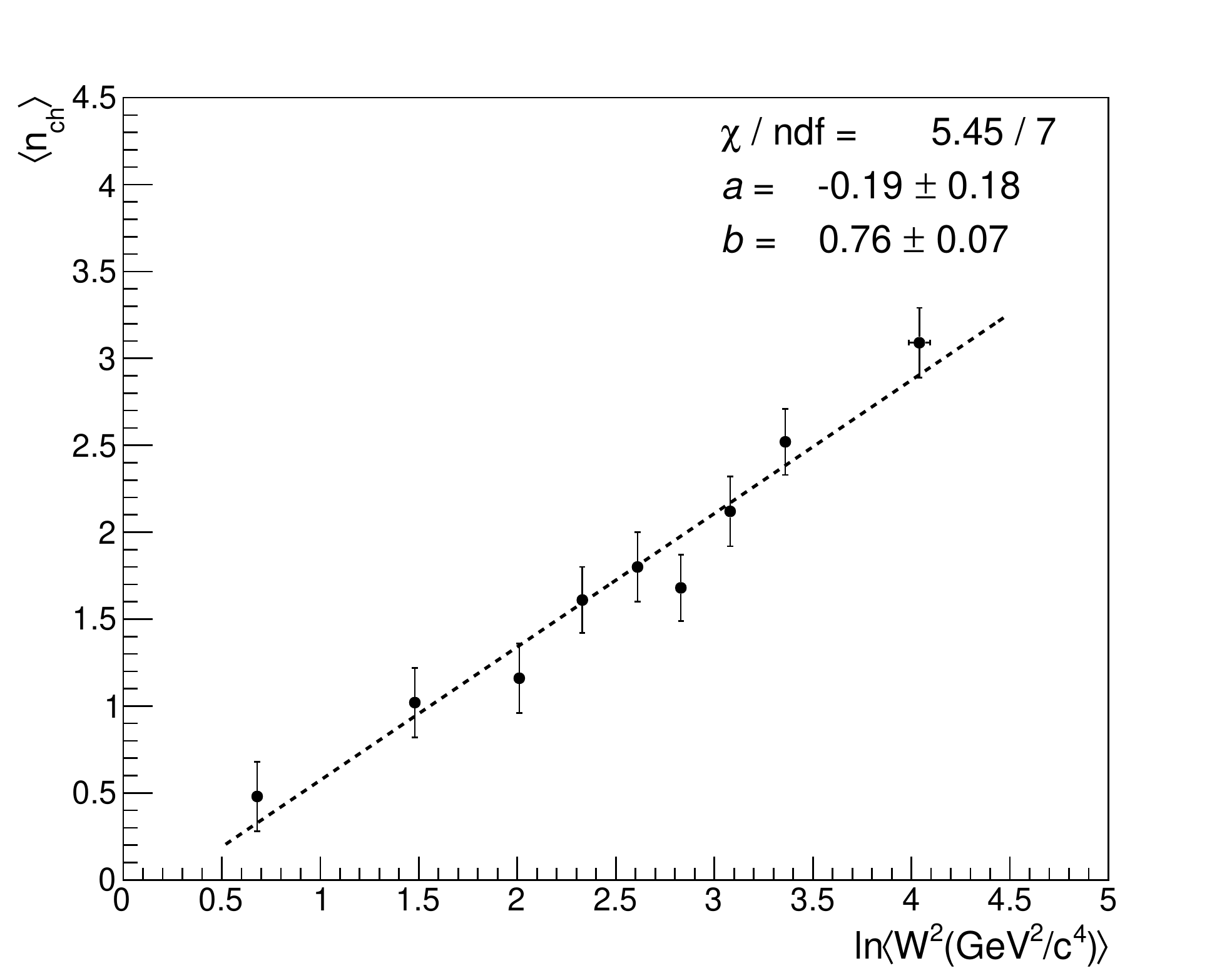}
  \caption{ The average charged hadron multiplicity distributions as a function of $lnW^{2}$.}\label{fig:lnw2}
\end{figure}

The average charged hadrons  multiplicity as a function of $W^{2}$ is 
presented in Figure~\ref{fig:lnw2}. The data is well described 
by a linear function in $lnW^2$:
\begin{linenomath*}
\begin{equation}
\langle n_{ch}\rangle = a + blnW^2
\end{equation}
\end{linenomath*}
The values of the fitted parameters are $a =  -0.19\pm 0.18$  and $ b = 0.76\pm 0.07$.
Values for different $W^2$ bin intervals are shown  in Table~\ref{tab:efftable}.  A comparison with other 
experiments is given in Table ~\ref{tab:comp}. 
\begin{table*}[ht]
\caption{The charged hadron multiplicity $n_{ch}$ distribution as a function of $W^2$(errors shown are statistical only).}\label{tab:efftable}
\label{sphericcase}
\begin{tabular*}{\textwidth}{@{\extracolsep{\fill}}lrrrrrrrrrrrrrrl@{}}
\hline
{$W^{2}(\mbox{GeV}^{2}/\mbox{c}^{4}$)}&\multicolumn{1}{c}{\small{ln$\langle W^{2}(\mbox{GeV}^{2}/\mbox{c}^{4})\rangle$}}& \multicolumn{1}{c}{\(\)}  & \multicolumn{1}{c}{\(\)} &
\multicolumn{1}{c}{} & \multicolumn{1}{c}{} &  \multicolumn{1}{c}{}  &  \multicolumn{1}{c}{} &\multicolumn{1}{c}{\(n_{ch}\)} &\multicolumn{1}{c}{}
&\multicolumn{1}{c}{} &\multicolumn{1}{c}{}&\multicolumn{1}{c}{} &\multicolumn{1}{c}{} &\multicolumn{1}{c}{} \\
\hline
{}& \multicolumn{1}{c}{} & \multicolumn{1}{c}{\(0\)} & \multicolumn{1}{c}{\(1\)} & \multicolumn{1}{c}{\(2\)} & \multicolumn{1}{c}{\(3\)} &
\multicolumn{1}{c}{\(4\)}  &  \multicolumn{1}{c}{\(5\)} &\multicolumn{1}{c}{\(6\)} &\multicolumn{1}{c}{\(7\)}  &\multicolumn{1}{c}{\(8\)}
&\multicolumn{1}{c}{\(9\)}&\multicolumn{1}{c}{\(10\)} &\multicolumn{1}{c}{\(\ge11\)} &$ \langle n_{ch}\rangle$ &\multicolumn{1}{c}{Total}  \\
\hline
1-3 & 0.68 \(\pm\)0.03 & 59  & 21 & 7 & 1& 0 & 1 & 0 & 0 & 0 &0 &0&0&0.48 \(\pm\) 0.11&89 \\
3-6 & 1.48 \(\pm\)0.01 & 29  & 37 & 24 & 4 & 2 & 0 & 1 & 0 & 0 &0 &0&0& 1.14 \(\pm\) 0.10&97 \\
6-9 & 2.01 \(\pm\)0.01 & 28  & 37 & 16 & 12 & 5 & 1& 1 & 0 & 0 &0 &0&0& 1.36 \(\pm\) 0.12&100 \\
9-12  & 2.33 \(\pm\)0.01 & 10  & 26 & 25 & 13 & 6 & 2 & 1 & 0 & 0 &0  &0&0&1.86 \(\pm\) 0.13&83 \\
12-15  & 2.61 \(\pm\)0.01 & 10  & 24 & 29 & 27 & 3 & 2 & 3 & 0 & 0 &0  &0&0&2.07 \(\pm\) 0.13&98 \\
15-19 & 2.83 \(\pm\)0.01 & 12  & 17 & 35 & 14 & 8 & 4 & 1 & 1 & 0 &0  &0&0&2.10 \(\pm\) 0.15 &92 \\
19-25 & 3.08 \(\pm\)0.01 & 5  & 18 & 22 & 19 & 12 & 3 & 1 & 3 & 0 &0  &0&0&2.51 \(\pm\) 0.16 &83 \\
25-35  & 3.36 \(\pm\)0.01 & 4  & 15 & 17 & 19 & 12 & 8 & 2 & 1 & 1 &0  &0&0&2.79 \(\pm\) 0.18&79 \\
 \(\ge\)35   & 4.04 \(\pm\)0.05 & 3  & 6 & 16 & 14 & 15 & 8 & 4 & 4 & 3&0 &1&0&3.59 \(\pm\) 0.24 &74 \\\hline
{Total}  & $2.83\pm 0.03$ & 160 & 201 & 191 & 123 & 63 & 29 & 14 & 9 & 4 & 0 &1 &0 & $1.94\pm 0.05$ &  795\\ 
\end{tabular*}
\end{table*}
\begin{table}[hb]
\centering
\caption{Values of the parameters of the linear fit to the average charged hadrons multiplicity dependence on $lnW^2$. The results from other
experiments are also shown for  comparison.}\label{tab:comp}
\label{parset}
\begin{tabular*}{\columnwidth}{@{\extracolsep{\fill}}lrrrr@{}}
\hline
{ Reaction} &  \multicolumn{1}{c}{$\langle E_{\nu}\rangle$(GeV)} &   \multicolumn{1}{c}{a} &  \multicolumn{1}{c}{b}  &   \multicolumn{1}{c}{Ref.}\\
\hline
 \(\nu_{\mu}\)-emulsion  &38 & 0.45\(\pm\) 0.24  & 0.94\(\pm\) 0.08 & \cite{CHORUS}\\
 \(\nu_{\mu}\)-emulsion & 50 & 1.92 \(\pm\)  0.68 & 1.19\(\pm\) 0.23 & \cite{Voyvodik}\\
 \(\nu_{\mu}\)-emulsion &8.7 & 1.07 \(\pm\)  0.05 & 1.32\(\pm\) 0.11 & \cite{Aleshin} \\
 \(\nu_{\mu}\)-lead & 20 & -0.19 \(\pm\)  0.18 & 0.76\(\pm\) 0.07 & OPERA\\
\end{tabular*}
\end{table}
\subsection{Dispersion}

\begin{figure}[!h]
  \centering
    \includegraphics[scale=0.45]{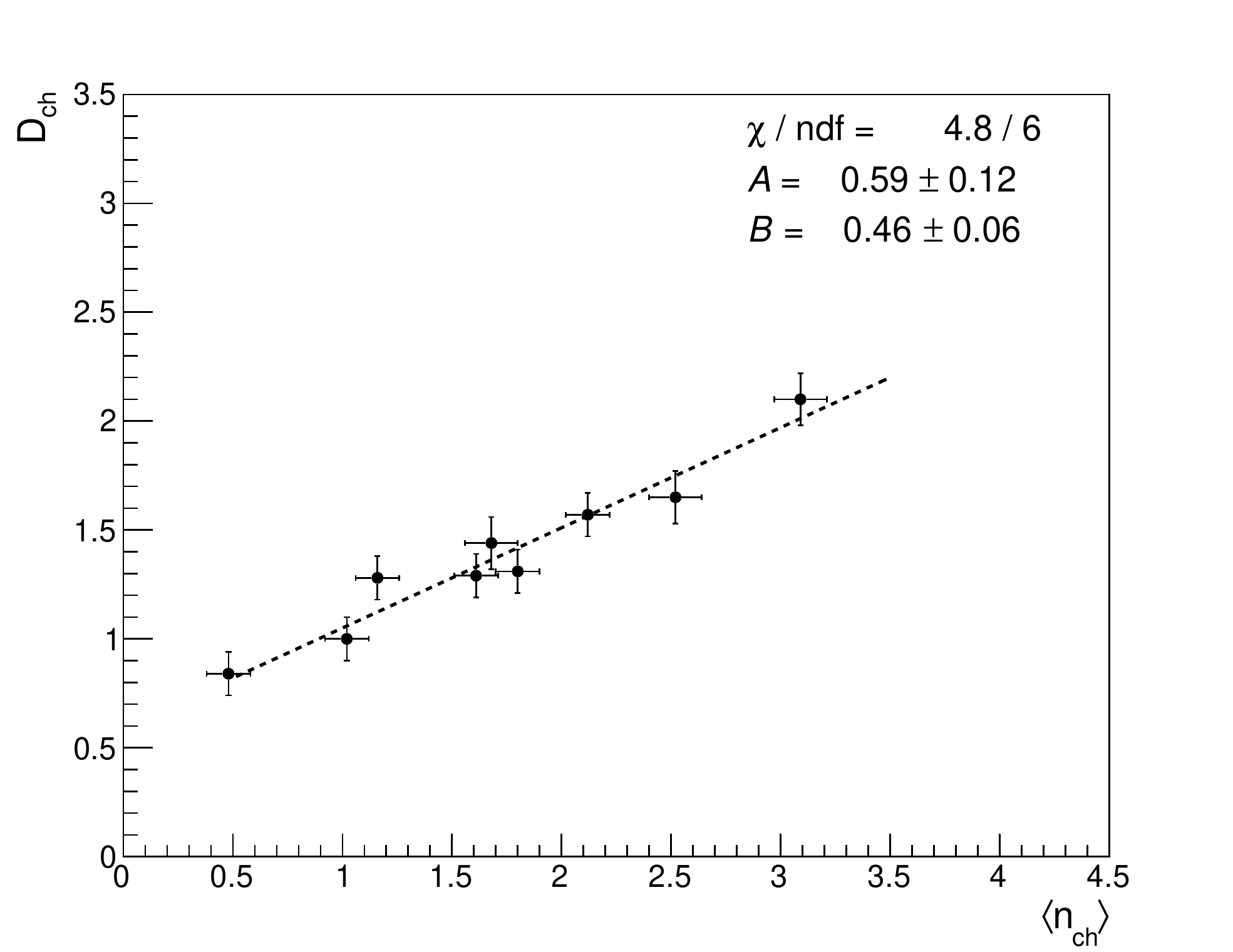}
  \caption{ The charged hadrons multiplicity dispersion as a function of \(\langle n_{ch}\rangle\).} \label{fig:disp}
\end{figure}

One of the characteristics of the multiplicity distribution which is of considerable theoretical interest is its dispersion.
In this section we investigate its dependence on the average multiplicity. The dispersion $D_{ch}$ is
defined as $D_{ch}$ $= \sqrt{\langle n^{2}_{ch}  \rangle - \langle n_{ch}  \rangle^{2}}$.
For independent particle production it  would follow a Poisson distribution with $D_{ch}$ $= \sqrt{\langle n_{ch} \rangle}$.
However, it was observed that charged particles production in hadronic interactions satisfies an  empirical parameterisation \cite{shivp}:
\begin{linenomath*}
\begin{equation}\label{eq:kno}
D_{ch} = A+B\langle n_{ch} \rangle
\end{equation}
\end{linenomath*}
Figure~\ref{fig:disp}  shows the dependence of the dispersion
on the average  multiplicity  \(\langle n_{ch}\rangle\) with a linear fit
superimposed.
The values of the fitted parameters are $A =  0.59\pm 0.12$ and $B = 0.46\pm 0.06$. They are shown in Table~\ref{tab:dch} together
with those obtained in other experiments for comparison.
\begin{table}[!h]
\centering
\caption{Values of parameters  $A$ and $B$ obtained by a linear  fit on the distribution of $D_{ch}$ versus  \(\langle n_{ch}\rangle\). The results
obtained in other neutrino experiments are also shown.}\label{tab:dch}
\centering
\label{parset}
\begin{tabular*}{\columnwidth}{@{\extracolsep{\fill}}lrrrr@{}}
\hline
{ Reaction} &  \multicolumn{1}{c}{$\langle E_{\nu}\rangle$(GeV)} & \multicolumn{1}{c}{$A$} &  \multicolumn{1}{c}{$B$}  &   \multicolumn{1}{c}{Ref.}\\
\hline
  \(\nu_{\mu}\)-emulsion  &38  & 1.18\(\pm\) 0.17  & 0.20\(\pm\) 0.05 & \cite{CHORUS}\\

 \(\nu_{\mu}\)-p & $>5$ & 0.36 \(\pm\)  0.03 & 0.36\(\pm\) 0.03 &  \cite{Allen}\\

   \(\nu_{\mu}\)-lead & 20  & 0.59 \(\pm\)  0.12 & 0.46\(\pm\) 0.06 & OPERA \\
\end{tabular*}
\end{table} 
\newpage
\subsection{KNO Scaling}

\begin{figure}[H]
  \centering
    \includegraphics[scale=0.45]{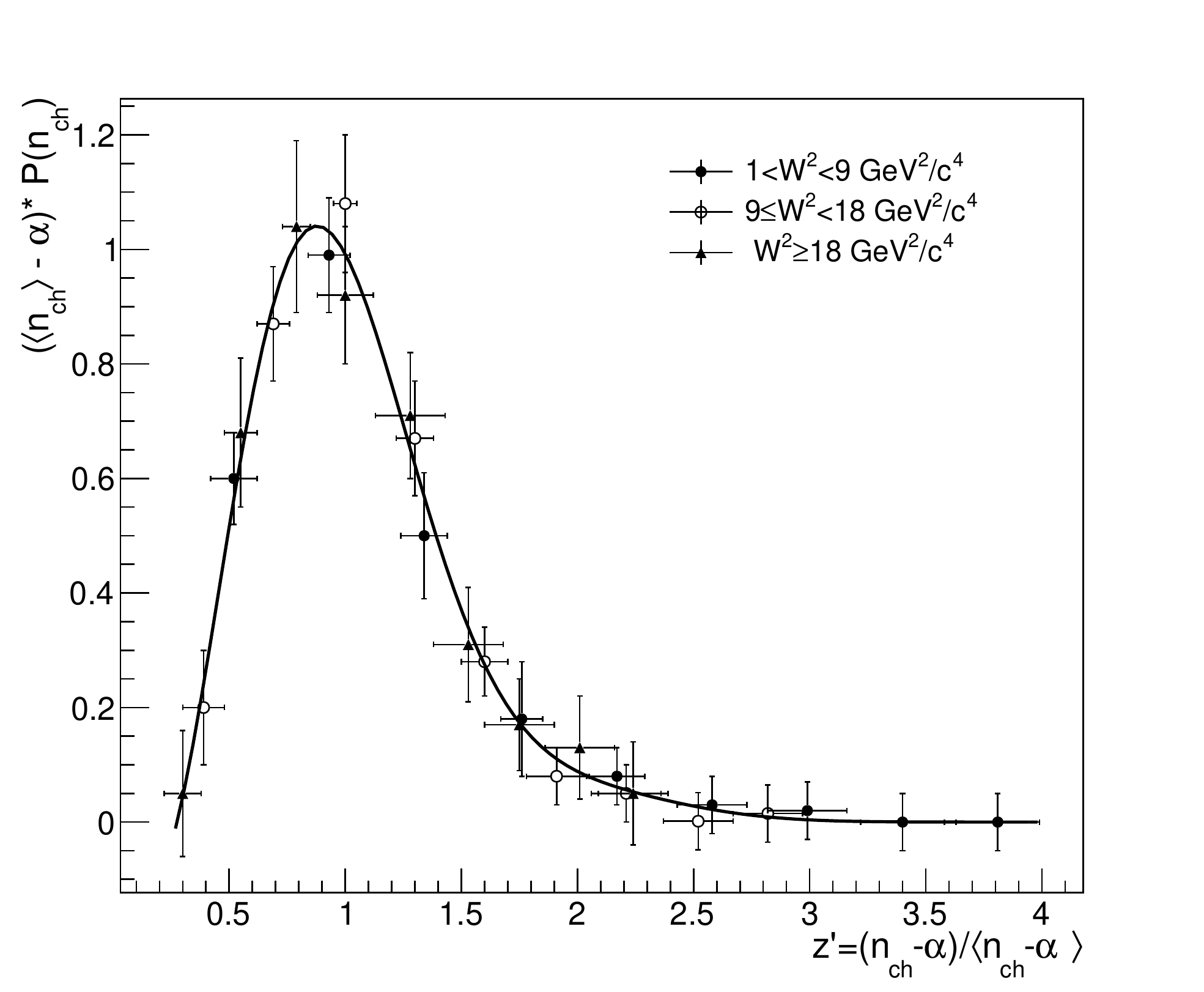}
  \caption{KNO scaling distribution. The curve superposed on the data results from a fit using the parameterisation defined in Eq. \ref{eq:func}.}\label{fig:kno}
\end{figure}

Koba, Nielsen, and Olesen have shown that  the multiplicity distribution $P(n_{ch})$ scaled by the average 
multiplicity \(\langle n_{ch}\rangle\) is asymptotically independent of the primary energy and depends only on  
variable 
\(z = {n_{ch}\over{\langle n_{ch} \rangle}}\):
\begin{linenomath*}
\begin{equation}
\langle n_{ch}  \rangle  \, .\,   P(n_{ch})\xrightarrow  {E \rightarrow \infty }
\Psi(z)
\end{equation}
\end{linenomath*}
KNO scaling is derived from Feynman scaling, i.e. 
based  on  the assumption that the rapidity density $dn_{ch}\over dy$ reaches its limit value at 
{\it y} = 0 above  a certain energy which corresponds to an asymptotic scaling of the total multiplicity
 as \({\langle n_{ch} \rangle}\propto ln\sqrt{s}\). KNO scaling implies that the intercept 
$A$ in Eq. \ref{eq:kno}  be 
compatible with 0, which is not the case at low to medium energies for all kinds of interactions. 
Buras et al. ~\cite{Buras} have introduced a new variable $z'$ defined as
\begin{linenomath*}
\begin{equation}
z' = \frac{ n_{ch} - \alpha} {\langle n_{ch}  - \alpha \rangle} 
\end{equation}
\end{linenomath*}
where the reaction dependent and energy independent parameter $\alpha$   is chosen in order 
to provide an extension of the KNO scaling to low energies. 
In the current analysis, this implies that $\alpha$ = $A\over{B}$ = 1.28. Hence
\begin{linenomath*}
\begin{equation}
\langle (n_{ch} - \alpha)\rangle P(n_{ch}) =\Psi(z')
\end{equation}
\end{linenomath*}
A tentative explanation for a non-zero value for $\alpha$ has been proposed in 
terms of a leading particle effect in interactions of 
hadrons~\cite{Stattery} and of neutrinos~\cite{baranov} as well as resulting from the heavy nuclear targets in neutrinos 
experiments using emulsion~\cite{CHORUS}. 
Figure~\ref{fig:kno} shows the distributions obtained for \(\Psi\) as a function of $z'$
for three different intervals of $W^2$ for \(\nu_\mu\)-lead CC interactions. 
The fitted curve superimposed on the data results from a parameterisation of \(\Psi (z')\)
\begin{linenomath*}
\begin{equation}\label{eq:func}
\Psi (z') = (Az' +B(z')^3 +C(z')^5 + D(z')^7)e^{-(Ez'+F(z')^3)} \\ 
\end{equation}
\end{linenomath*}
The fit parameters are given in Table ~\ref{tab:fit}. 
The data shows good agreement with approximate KNO scaling. Such observation has been made by other experiments \cite{CHORUS}, \cite{hebert,hebert2,wroblewski,baranov}.
\begin{table}[h!]
\caption{The best-fitted parameters of the function given by Eq. \ref{eq:func}.}\label{tab:fit}
\centering
\label{parset}
\begin{tabular}{c|c}
$A$ &~~~ $-1.6\pm{2.4}$\\
$B$ &~~~ $22.7\pm{18.8}$\\
$C$ &~~~ $-9.4\pm{17.1}$\\
$D$ &~~~ $1.4\pm{0.9}$ \\
$E$ &~~~ $2.3\pm{1.2}$\\
$F$ &~~~ $-0.2\pm{-0.2}$ \\ 
\end{tabular}
\end{table}

\section{Conclusion}

In this article, we studied the characteristics of the multiplicity distribution of charged hadrons in neutrino-lead interactions 
in the OPERA detector with the main objective to 
aid in tuning the models used in MC event generators. 
For this purpose, the results are presented in detail in the form of tables. 
They can be summarized as follow:
\begin{enumerate}[i]
\item The dependence of  the average multiplicity $\langle n_{ch} \rangle$ on $ln~W^2$  is approximately linear.

\item The dependence on the  charged hadrons multiplicity $n_{ch}$ of its dispersion $D_{ch}$ is approximately linear.
\item Approximate KNO scaling is valid for the
charged hadrons multiplicity.
\end{enumerate}

\section*{Acknowledgments}

We acknowledge CERN for the successful operation of the CNGS facility and INFN for the continuous 
support given to the experiment through its LNGS laboratory. We acknowledge funding from our national 
agencies: Fonds de la Recherche Scientifique-FNRS and Institut Interuniversitaire des Sciences 
Nucleaires for Belgium; MoSES for Croatia; CNRS and IN2P3 for France; BMBF for Germany; INFN for 
Italy; JSPS, MEXT, the QFPU-Global COE program of Nagoya University, and Promotion and Mutual 
Aid Corporation for Private Schools of Japan for Japan; SNF, the University of Bern and ETH Zurich 
for Switzerland; the Russian Foundation for Basic Research (Grant No. 12-02-12142 ofim), the Programs 
of the Presidium of the Russian Academy of Sciences (Neutrino Physics and Experimental and Theoretical 
Researches of Fundamental Interactions), and the Ministry of Education and Science of the Russian Federation 
for Russia, the Basic Science Research Program through the National Research Foundation of Korea (NRF) funded by the Ministry of Science, ICT and Future Planning (Grant No. NRF-2015R1A2A 2A01004532) for Korea; and 
TUBITAK, the Scientific and Technological Research Council of Turkey for Turkey (Grant No. 108T324). 
We thank the IN2P3 Computing Centre (CC-IN2P3) for providing computing resources.

\end{document}